\documentclass[12pt,leqno]{amsart}
\usepackage{graphicx}
\usepackage[letterpaper,margin=1in]{geometry}
\usepackage{indentfirst,csquotes,setspace}

\usepackage{amssymb,amsthm,amsmath}
\usepackage{xcolor,paralist,titlesec,fancyhdr,etoolbox}

\usepackage[ruled,linesnumbered]{algorithm2e}
\SetAlgoNlRelativeSize{0}
\usepackage{subfigure}
\usepackage{enumerate}
\usepackage{booktabs}
\usepackage{threeparttable}
\usepackage{url}
\usepackage{multirow}
\usepackage[authoryear,round]{natbib}
\setcitestyle{authoryear,round,aysep={,},citesep={;}}
\usepackage{float}
\usepackage{placeins}
\usepackage{hyperref}

\theoremstyle{plain}
\newtheorem{theorem}{Theorem}

\newtheorem{proposition}{Proposition}
\newtheorem{corollary}{Corollary}

\newtheorem{assumption}{Assumption}
\theoremstyle{definition}
\newtheorem{definition}{Definition}
\newtheorem{example}{Example}
\theoremstyle{remark}
\newtheorem{remark}{Remark}

\titleformat{\section}[hang]
  {\normalfont\Large\bfseries}{\thesection}{0.8em}{}
\titlespacing*{\section}{0pt}{3.5ex plus 1ex minus .2ex}{2ex plus .2ex}
\titleformat{\subsection}[hang]
  {\normalfont\bfseries}{\thesubsection}{0.8em}{}
\titlespacing*{\subsection}{0pt}{2.5ex plus .8ex minus .2ex}{1ex plus .2ex}
\titleformat{\subsubsection}[hang]
  {\normalfont\itshape}{\thesubsubsection}{0.8em}{}
\titlespacing*{\subsubsection}{0pt}{2ex plus .5ex minus .2ex}{.8ex plus .2ex}
\makeatletter
\renewcommand{\sectionmark}[1]{%
  \begingroup
  \edef\x{\endgroup
    \noexpand\markboth
      {\number\c@section\quad\noexpand\MakeUppercase{\unexpanded{#1}}}%
      {\number\c@section\quad\noexpand\MakeUppercase{\unexpanded{#1}}}}%
  \x}
\providecommand{\@secnumpunct}{.\quad}
\makeatother

\makeatletter
\patchcmd{\@settitle}{\uppercasenonmath\@title}{}{}{}
\patchcmd{\@setauthors}{\MakeUppercase{\authors}}{\authors}{}{}
\makeatother

\hypersetup{hidelinks}

\makeatletter
\newcommand{\DeclareSupplementRef}[2]{%
	\expandafter\def\csname r@#1\endcsname{{#2}{0}}}
\DeclareSupplementRef{Appendix A}{A}
\DeclareSupplementRef{Appendix C}{C}
\DeclareSupplementRef{app:exhaustive-test}{C.1}
\DeclareSupplementRef{app:exhaustive-simultaneous}{C.3}
\DeclareSupplementRef{table:Variables in New York taxi data}{S4}
\makeatother

\newcommand{\indep}{\perp\!\!\!\perp}
\newcommand{\dep}{\not \! \perp\!\!\!\perp}
\newcommand{\spacingset}[1]{\renewcommand{\baselinestretch}{#1}\small\normalsize}

\begin{document}
\setstretch{1.7}
\title{Marginal Coordinate Test for Fréchet Regression with Random Objects}
\author[J. Chen, R. Qiu, R. Wang, and Z. Yu]{%
Jiaye Chen$^{1}$, Rui Qiu$^{1}$, Roulin Wang$^{2}$, and Zhou Yu$^{1}$\\
\textnormal{\footnotesize $^{1}$School of Statistics, East China Normal University}\\
\textnormal{\footnotesize $^{2}$School of Management, University of Science and Technology of China}}
\date{}

\begin{abstract}
We develop a marginal coordinate test for regression with Euclidean predictors and a random-object response in a separable metric space. The goal is to test whether a predictor provides additional information about the response conditional on the remaining predictors. In a semi-supervised design, an unlabeled sample is used to estimate predictor conditional means, while an independent labeled sample is reserved for inference. The resulting residuals are combined with a product-space kernel to form a kernel conditional mean dependence (KCMD) U-statistic without requiring a response residual. The primary identity-based test targets a necessary conditional mean restriction, while a multiple-transformation extension probes broader alternatives. We establish a weighted centered chi-square null limit, wild bootstrap validity, consistency against fixed detectable alternatives, and local power under mean-element alternatives. For simultaneous inference, truncated p-to-e calibration combined with e-BH provides asymptotic false discovery rate control under general dependence. Simulations with Euclidean and non-Euclidean responses, together with a New York City taxi-flow analysis, illustrate the method.
\end{abstract}


\maketitle

\bigskip

	\section{Introduction}
	\label{sec:intro}
	Regression responses are increasingly observed as probability distributions, symmetric positive definite matrices, shapes, networks, and other random objects whose geometry is more naturally described by a metric than by Euclidean coordinates. Fr\'{e}chet regression provides a general framework for studying their conditional mean \citep{petersen2019frechet}, with subsequent developments such as single-index modeling \citep{bhattacharjee2023single}, the random forest method \citep{qiu2024random} and deep learning \citep{iao2025deep}. Sufficient dimension reduction has also been extended to Fr\'{e}chet regression through weighted inverse regression ensembles \citep{ying2022frechet} and ensembles of response transformations \citep{zhang2024dimension}, followed by work on variable selection \citep{tucker2023variable} and feature screening \citep{tian2025feature}.
	These advances provide useful tools for estimation and dimension reduction with metric space responses. They do not, however, directly provide a calibrated marginal coordinate test for whether an original predictor remains informative after the others have been taken into account.
	
	This paper addresses that inferential question. Let $X=(X_1,\ldots,X_p)$ be a Euclidean predictor and let $Y$ take values in a separable metric space. For each coordinate, we consider
	$$H_{0j}:Y\indep X_j \mid X_{-j}
	\quad\text{versus}\quad
	H_{1j}:Y\dep X_j \mid X_{-j}.$$
	The hypothesis concerns the contribution of $X_j$ to the full conditional law of the random-object response, rather than to a specified conditional mean. It is neither a test of marginal association nor a screening criterion, because the remaining predictors are explicitly retained. Testing these hypotheses separately or simultaneously yields coordinatewise inference on predictor contribution without requiring estimation of an entire sufficient subspace.
	
	The origin of this target lies in sufficient dimension reduction. A sufficient reduction seeks a matrix $B$ such that $Y\indep X\mid B^TX$, and the central subspace $\mathcal{S}_{Y\mid X}$ is the smallest subspace with this property when it exists \citep{li1991sliced}. For $\mathcal{H}_j=\operatorname{span}(e_j)$, \citet{cook2004testing} showed that the marginal coordinate hypothesis $P_{\mathcal{H}_j}\mathcal{S}_{Y\mid X}=\{0\}$ is equivalent to $H_{0j}$. Classical tests assess this hypothesis through method-specific inverse regression information, including sliced inverse regression \citep{cook2004testing}, sliced average variance estimation \citep{shao2007marginal}, and directional regression \citep{yu2016model}. \citet{dong2016permutation} developed permutation calibration for these procedures. More recent work has incorporated simultaneous false discovery rate (FDR) control \citep{guo2024model} and multivariate responses \citep{dong2026marginal}. These procedures show that coordinate tests inherit the coverage properties of the corresponding sufficient dimension reduction methods used to construct them. Their statistics and calibration, however, are designed around scalar or Euclidean response representations, whereas existing Fr\'{e}chet sufficient dimension reduction methods primarily estimate subspaces or rank predictors. Consequently, they do not directly deliver a calibrated marginal coordinate test for a general metric space response.
	
	Viewed probabilistically, Cook’s marginal coordinate hypothesis is a conditional independence hypothesis. Existing approaches include kernel conditional covariance \citep{fukumizu2008kernel}, conditional distance correlation \citep{wang2015conditional}, and residual constructions \citep{lundborg2022conditional}. These methods are not directly tailored to the present setting: conditional distance correlation is formulated for multivariate Euclidean variables, residual methods commonly rely on Euclidean or Hilbert-space response regressions, and kernel methods for general domains require additional conditional operator estimation. A useful construction should instead exploit the response metric while avoiding an ordinary response residual.
	
	Our main idea is to remove from transformations of $X_j$ the part explained by $X_{-j}$ and then assess the remaining conditional association with the pair $(Y, X_{-j})$. Over the complete class of integrable measurable transformations, the resulting conditional moment restrictions characterize Cook's marginal coordinate hypothesis exactly. A characteristic kernel on the product space converts these restrictions into scalar criteria, while pairwise response distances carry the geometry of $Y$. Retaining $X_{-j}$ in this product space is essential: it preserves the conditional nature of the hypothesis instead of averaging the conditioning variables out. The construction therefore provides a metric-compatible route to the same coordinate hypothesis studied in sufficient dimension reduction, without estimating either the structural dimension or a response-side regression.
	
	The complete transformation class is a population benchmark rather than a literal implementation. The primary Fr\'{e}chet marginal coordinate test (FMCT) uses the identity transformation, which is stable and computationally economical but may miss contributions expressed only through nonlinear features of a predictor. A finite collection of transformations can probe a broader range of alternatives at the cost of additional nuisance estimation. This coverage and complexity tradeoff parallels the role of exhaustiveness in sufficient dimension reduction: richer information can improve detection, but a finite family recovers the complete active set only under an appropriate exhaustiveness condition.
	
	For sample inference, the required nuisance regressions involve only Euclidean predictors. This permits a semi-supervised implementation in which an unlabeled predictor sample is used for nuisance estimation and an independent labeled sample is used to construct the test statistic. We derive the asymptotic null distribution and establish wild bootstrap calibration. For simultaneous marginal coordinate testing, we convert the resulting p-values into truncated e-values and apply e-BH \citep{wang2022false}, obtaining asymptotic FDR control under general dependence among the coordinate hypotheses.
	
	Our contributions are threefold.
	First, we give an exact transformation-indexed characterization of Cook's coordinate hypothesis for a metric space response and show that finite families recover the active set under coordinate exhaustiveness. The product-space criterion retains both the response geometry and the conditioning variables without requiring central subspace estimation or a Euclidean response representation.
	Second, we develop a semi-supervised FMCT that accommodates flexible predictor-side nuisance estimation, including deep neural networks, and establish its asymptotic null distribution, wild bootstrap validity, and power under fixed and local detectable alternatives.
	Third, we combine coordinatewise bootstrap p-values with truncated p-to-e calibration and e-BH to obtain asymptotic FDR control under arbitrary dependence.
	
	We organize the rest of this paper as follows. 
	Section~\ref{sec:reformulate} develops population criteria for the marginal coordinate hypothesis. Section~\ref{sec:Test} constructs the semi-supervised statistic and develops its null, bootstrap, and power theory. Section~\ref{sec:FDR} extends the coordinatewise tests to simultaneous inference through e-values. Sections~\ref{sec:Simulation} and~\ref{sec:Real data analysis} report the numerical experiments and the taxi flow analysis, respectively, and Section~\ref{Conclusion} concludes. Data details, proofs, and the multiple transformation procedure are collected in the Supplementary Material.
    
	\section{Population characterization}\nopagebreak
	\label{sec:reformulate}
	This section develops the population foundation of FMCT in three steps. We first give an exact characterization of the marginal coordinate hypothesis using integrable transformations. We then specify the conditions under which a finite transformation family can recover the active coordinates. Finally, we introduce the identity transformation used by the primary test and clarify the information that this simpler specification retains.
	
	\subsection{Exact population criterion}
	\label{sec:primary-population-criterion}
	Let $X=(X_1,...,X_p)^\top\in \mathbb{R}^p$, and let $X_j$ denote its $j$th component. Define
	$$X_{-j}=(X_1,...,X_{j-1},X_{j+1},...,X_p)^\top.$$
	The response $Y$ takes values in a separable metric space $(\mathcal{Y},d)$. Fix a coordinate $j$ and recall that $H_{0j}$ states $Y\indep X_j\mid X_{-j}$. In words, after $X_{-j}$ has been fixed, observing $Y$ should not change the conditional behavior of $X_j$. Define the class
	\[
	\mathcal G_j
	=
	\left\{
	g:\mathbb R\to\mathbb R:
	g\text{ is Borel measurable and }E|g(X_j)|<\infty
	\right\}.
	\]
	For $g\in\mathcal G_j$, define
	$$V_j^g=g(X_j)-E\{g(X_j)\mid X_{-j}\}.$$
	By construction, $E(V_j^g\mid X_{-j})=0$. Under $H_{0j}$, observing $Y$ cannot alter the conditional average of $g(X_j)$, and hence $E(V_j^g\mid Y,X_{-j})=0$. Retaining $X_{-j}$ in this conditioning set is essential. The weaker restriction $E(V_j^g\mid Y)=0$ averages over $X_{-j}$ and can miss departures that vary with the remaining predictors. Conversely, if the restriction conditional on $(Y, X_{-j})$ holds for every $g\in\mathcal G_j$, it holds in particular for all indicator functions of $X_j$; the resulting conditional probabilities determine the conditional distribution of $X_j$, giving conditional independence. Thus, the full integrable transformation class yields an exact population characterization, whereas a fixed $g$ examines only one aspect of the conditional behavior of $X_j$. Here, residualization in $V_j^g$ removes the part of $g(X_j)$ explained by the remaining predictors and does so entirely on the Euclidean predictor side, without defining a residual for $Y$.
	
	For a fixed $g$, we next seek a scalar representation of the conditional moment restriction $E(V_j^g\mid Y, X_{-j})=0$. Because the conditioning variables include the metric space response, we first represent its geometry in a Hilbert space. Following \citet{LYONS}, we impose the following condition.
	
	\begin{assumption}\label{assumption:negative}
		$(\mathcal{Y} ,d)$ is of negative type.
	\end{assumption}
	Under Assumption~\ref{assumption:negative}, there exist a separable Hilbert space $\mathcal{H}$ and a map $\phi : \mathcal{Y}\rightarrow \mathcal{H}$ such that, for all $Y,Y^{\prime}\in\mathcal{Y}$,
	$$d_{\mathcal{Y}}(Y , Y^{\prime})=\| \phi (Y)-\phi (Y^{\prime})\Vert_{\mathcal{H}}^2 .$$
	Examples of negative-type metric spaces include Euclidean spaces with
	their usual distance, univariate probability distributions with the
	2-Wasserstein distance, symmetric positive-definite (SPD) matrices with the Log-Cholesky distance, and
	unit spheres with geodesic distance. These are precisely the response
	geometries considered in our numerical experiments; see
	\citet{schoenberg1938metric,LYONS} for these and further examples.
	
	Accordingly, for the $j$th coordinate, define
	$\Gamma_j=(\phi(Y),X_{-j})$. The random element $\Gamma_j$ represents the
	full conditioning information $(Y,X_{-j})$ in a Hilbert product space.
	
	\begin{assumption}[Kernel regularity]\label{assump:bounded-kernel}
		The kernel $K$ is measurable, positive definite, and characteristic on the support of
		$\Gamma_j$, and its reproducing kernel Hilbert space (RKHS) $\mathcal H_K$ is separable. In addition, $\sup_{\gamma}K(\gamma,\gamma)\leq\kappa^2$ for some $\kappa<\infty$.
	\end{assumption}
	
	The kernel-based conditional mean dependence (KCMD) provides a nonnegative scalar measure of a conditional mean restriction. In general, let $T$ and $W$ be random elements in separable Hilbert spaces $\mathcal{T}$ and $\mathcal{W}$, respectively, with $E\|W\|_{\mathcal W}<\infty$. Let
	$(\tilde{T},\tilde{W})$ be an independent copy of $(T,W)$.
	For a kernel satisfying the corresponding conditions in
	Assumption~\ref{assump:bounded-kernel} on the support of $T$, define
	$$\text{KCMD}(W,T)=E \{K(T,\tilde{T}) \langle W-EW,\tilde{W}-E\tilde{W}\rangle_{\mathcal{W}}  \}.$$
	Proposition 1 of \citet{lai2021kernel} gives the following properties:
	\begin{itemize}
		\item[(i)]  $\text{KCMD}(W,T) \geq 0$;
		\item[(ii)]  $\text{KCMD}(W,T)=0
		\ \text{if and only if} \ E(W \mid T)=E(W)\ \text{a.s.}$.
	\end{itemize}
	Under Assumptions~\ref{assumption:negative} and
	\ref{assump:bounded-kernel}, take $W=V_j^g$ and $T=\Gamma_j$. Since
	$E(V_j^g)=0$, property (ii) shows that $\text{KCMD}(V_j^g,\Gamma_j)=0$ is
	equivalent to $E(V_j^g\mid Y,X_{-j})=0$. Combining this equivalence
	with the preceding transformation argument yields the following characterization.
	\begin{theorem}\label{thm:population-necessity}
		Fix $j\in\{1,\ldots,p\}$. Then $H_{0j}$ holds if and only if, for every
		$g\in\mathcal G_j$,
		\[
		E(V_j^g\mid Y,X_{-j})=0 \quad \text{a.s.}
		\]
		If, in addition,
		Assumptions~\ref{assumption:negative} and
		\ref{assump:bounded-kernel} hold, then they are
		also equivalent to
		\[
		\Lambda_j^g:= \text{KCMD} (V_j^g,\Gamma_j)=0
		\quad\text{for every }g\in\mathcal G_j.
		\]
	\end{theorem}
	Theorem~\ref{thm:population-necessity} separates the exact population characterization from the specifications used in practice. A positive $\Lambda_j^g$ for any transformation refutes $H_{0j}$, but a zero value for one selected transformation is not conclusive. To recover the converse, the chosen transformations must collectively retain enough information about the conditional behavior of $X_j$. Sections~\ref{sec:exhaustive-extension} and~\ref{sec:identity-specification} make this distinction precise for a finite family and for the identity transformation, respectively. None of these population criteria requires estimation of the central subspace, its structural dimension, or a response-side regression.
	\begin{remark}\label{rem:product-space-kernel}
		Although $\Gamma_j$ is defined through $\phi$, the embedding need not be constructed explicitly when using a radial or distance-induced kernel. The kernel is defined on $(\mathcal{H} \times \mathbb{R}^{p-1}) \times (\mathcal{H} \times \mathbb{R}^{p-1})$. The product $\mathcal{H} \times \mathbb{R}^{p-1}$ remains a separable Hilbert space under the
		inner product $\langle (a,b),(a',b')\rangle_{\mathcal{H} \times \mathbb{R}^{p-1}} = \langle a,a' \rangle_{\mathcal{H}} + \langle b,b' \rangle_{\mathbb{R}^{p-1}}$
		for $a,a' \in \mathcal{H}$ and $b,b' \in \mathbb{R}^{p-1}$,
		so the KCMD construction is well defined in the product space. Let $\tilde\Gamma_j=(\phi(\tilde Y),\tilde X_{-j})$ be an independent copy of $\Gamma_j$. By the defining property of $\phi$,
		$$d_{\mathcal{H} \times \mathbb{R}^{p-1}}^2(\Gamma_j,\tilde\Gamma_j)
		=\|\phi(Y)-\phi(\tilde Y)\|_{\mathcal H}^2+\|X_{-j}-\tilde X_{-j}\|_2^2
		=d_{\mathcal Y}(Y,\tilde Y)+\|X_{-j}-\tilde X_{-j}\|_2^2.$$
		Hence, a radial or distance-induced kernel $K(\Gamma_j,\tilde\Gamma_j)$ can be evaluated directly from the observed response distance $d_{\mathcal Y}(Y,\tilde Y)$ and the Euclidean distance between $X_{-j}$ and $\tilde X_{-j}$.
	\end{remark}
	
	\subsection{Finite families and active set recovery}
	\label{sec:exhaustive-extension}
	The complete transformation class is a population benchmark rather than a literal implementation. To describe what a finite family must retain, define the active set
	$$\mathcal{A}=\{j:Y\dep X_j\mid X_{-j}\}\subseteq\{1,\ldots,p\}.$$
	In sufficient dimension reduction, $\mathcal A$ records the original coordinate directions that contribute to the central subspace. Under Assumptions~\ref{assumption:negative} and \ref{assump:bounded-kernel}, KCMD is nonnegative and Theorem~\ref{thm:population-necessity} gives the exact representation
	$$\mathcal A=\{j:\Lambda_j^g>0\text{ for at least one }g\in\mathcal G_j\}.$$
	
	Now select a common finite family $\mathcal G^{(m)}=\{g_1,\ldots,g_m\}$ of admissible transformations, with $g_q\in\mathcal G_j$ for every coordinate under consideration. For each coordinate, define the criterion vector
	$$\boldsymbol{\Lambda}_j(\mathcal G^{(m)})
	=\left(\Lambda_j^{g_1},\ldots,\Lambda_j^{g_m}\right)^T,$$
	and collect these vectors as the rows of $\Upsilon(\mathcal G^{(m)})=(\Lambda_j^{g_q})_{p\times m}$. Every inactive coordinate has a zero row, but an active coordinate can also have a zero row if none of the selected transformations reveals its contribution. Accordingly, the finite family detects
	$$\mathcal A(\mathcal G^{(m)})
	=\left\{j:\sum_{q=1}^m\Lambda_j^{g_q}>0\right\}
	\subseteq\mathcal A.$$
	This inclusion shows the population tradeoff precisely: restricting the transformation class cannot create a false active coordinate, but it can lose an active one.
	
	\begin{assumption}[Coordinate exhaustiveness]\label{assump:coordinate-exhaustiveness}
		The finite family $\mathcal G^{(m)}$ is coordinate exhaustive, in the sense that $\mathcal A(\mathcal G^{(m)})=\mathcal A$.
	\end{assumption}
	Equivalently, for every $j\in\mathcal A$, at least one $g_q\in\mathcal G^{(m)}$ satisfies $\Lambda_j^{g_q}>0$. This assumption concerns coverage relative to the underlying distribution; no fixed finite family is automatically exhaustive over all possible alternatives.
	
	\begin{remark}\label{rem:coordinate-exhaustiveness}
		The terminology is analogous to coverage and exhaustiveness conditions in sufficient dimension reduction \citep{yin2011sufficient} and related coordinate testing procedures \citep{dong2016permutation,guo2024model}. The present condition is stated directly in terms of row support because the inferential target is the active coordinate set. It is weaker than requiring the columns of $\Upsilon(\mathcal G^{(m)})$ to span the entire active coordinate subspace, which would impose rank $|\mathcal A|$ and hence require $m\geq|\mathcal A|$. A single transformation may detect several active coordinates.
	\end{remark}
	
	Assumption~\ref{assump:coordinate-exhaustiveness} restores the equivalence lost when the complete class is replaced by a finite family.
	
	\begin{proposition}\label{prop:active-set-characterization}
		Suppose that Assumptions~\ref{assumption:negative},
		\ref{assump:bounded-kernel}, and
		\ref{assump:coordinate-exhaustiveness} hold. For every
		$j\in\{1,\ldots,p\}$,
		$$H_{0j}\quad\Longleftrightarrow\quad
		\boldsymbol{\Lambda}_j(\mathcal G^{(m)})=0
		\quad\Longleftrightarrow\quad
		\sum_{q=1}^m\Lambda_j^{g_q}=0.$$
	\end{proposition}
	The first equivalence follows from active set recovery, and the second uses the nonnegativity of KCMD. Proposition~\ref{prop:active-set-characterization} motivates the aggregate hypothesis
	$$H_{0j}^{\prime\prime}:\sum_{q=1}^m\Lambda_j^{g_q}=0,
	\quad \text{versus}\quad
	H_{1j}^{\prime\prime}:\sum_{q=1}^m\Lambda_j^{g_q}\neq0.$$
	Under coordinate exhaustiveness, this is equivalent to the scientific hypothesis $H_{0j}$; without it, the aggregate remains a valid but potentially nonexhaustive necessary condition. Candidate families include polynomials, slicing indicators, the sine and cosine components of characteristic functions, and normalized B-spline bases \citep{guo2024model}. Supplementary Material~\ref*{Appendix C} develops inference based on multiple transformations, covering the test statistic, wild bootstrap calibration, asymptotic theory, and sensitivity analyses.
	
	\subsection{Primary identity specification}
	\label{sec:identity-specification}
	The finite-family formulation allows many choices of $g$, but the primary procedure requires a prespecified and interpretable baseline. Assume that $E|X_j|<\infty$ for each coordinate under consideration. Then the identity transformation $g(x)=x$ belongs to $\mathcal G_j$. This choice follows the inverse regression tradition represented by sliced inverse regression (SIR) and marginal coordinate tests derived from it, which use conditional means of predictors as their basic source of information \citep{li1991sliced,cook2004testing}. The connection is conceptual rather than an equivalence: FMCT residualizes one coordinate against $X_{-j}$ and measures its conditional mean dependence on $(Y,X_{-j})$ through a product space kernel, without slicing $Y$ or estimating an SIR matrix. Among single transformations, the identity retains a direct interpretation on the original predictor scale, avoids an additional transformation choice, and requires only the nuisance regression $E(X_j\mid X_{-j})$. We therefore adopt it as the primary specification. Write
	$$V_j=X_j-E(X_j\mid X_{-j}),\qquad
	\Lambda_j=\text{KCMD}(V_j,\Gamma_j),$$
	and consider the auxiliary population hypothesis
	$$H_{0j}^{\prime}:\Lambda_j=0,
	\quad \text{versus}\quad
	H_{1j}^{\prime}:\Lambda_j\neq0.$$
	Theorem~\ref{thm:population-necessity} ensures that $H_{0j}$ implies $H_{0j}^{\prime}$, so rejection of the auxiliary null is evidence against the marginal coordinate hypothesis. Unless the identity transformation is itself coordinate exhaustive, however, failure to reject $H_{0j}^{\prime}$ does not establish conditional independence.
	
	The identity specification is computationally economical and less exposed to accumulated nuisance estimation error than a collection of transformations. Its limitation is equally clear: it probes changes in the conditional mean of $X_j$ after $(Y,X_{-j})$ is observed and may miss contributions expressed only through nonlinear transformations or other features of the conditional distribution. For example, after $X_{-j}$ is held fixed, the conditional variance of $X_j$ may vary with $Y$ while its conditional mean remains unchanged. The primary FMCT should therefore be viewed as a parsimonious first moment procedure rather than an omnibus test. 

	\section{Semi-supervised coordinate inference}
	\label{sec:Test}
	This section turns the identity specification in Section~\ref{sec:identity-specification} into a coordinatewise testing procedure. An unlabeled sample is used to fit the nuisance regression, and an independent labeled sample is used to construct the pairwise statistic. We define the statistic, derive its asymptotic null distribution, establish wild bootstrap calibration, and analyze power under fixed and local alternatives. The extension to the finite-family procedure in Section~\ref{sec:exhaustive-extension} is collected in Supplementary Material~\ref*{Appendix C}.
	
	\subsection{Primary test statistic}
	\label{sec:semi-supervised-estimation}
	The scientific target remains Cook's marginal coordinate hypothesis $H_{0j}$. The primary statistic is constructed for the auxiliary criterion $H_{0j}^{\prime}$ in Section~\ref{sec:identity-specification}; because $H_{0j}$ implies $H_{0j}^{\prime}$, rejection of the latter also refutes the marginal coordinate null.
	
	Define the nuisance regression function $f_j(x_{-j})=E(X_j\mid X_{-j}=x_{-j}),$
	which is determined entirely by the predictor distribution $P_X$. Let
	$\mathcal{D}_1=\{X^{(i)}:1\leq i\leq n_1\}$
	be an i.i.d. predictor-only sample from $P_X$. Independently, let
	$\mathcal{D}_2=\{(X^{(k)},Y^{(k)}):1\leq k\leq n_2\}$
	be an i.i.d. labeled sample from the joint distribution of $(X,Y)$. Using $\mathcal D_1$, fit an estimator $\widehat f_{n_1,j}$ of $f_j$. For each labeled observation, define the estimated residual and pairwise kernel value by
	$$\widehat V_j^{(k)}=X_j^{(k)}-\widehat f_{n_1,j}(X_{-j}^{(k)}),
	\qquad
	K_{j,kl}=K(\Gamma_j^{(k)},\Gamma_j^{(l)}).$$
	The sample analogue of $\Lambda_j$ is
	$$T_{(n_1,n_2),j}
	=\frac{1}{n_2(n_2-1)}
	\sum_{1\leq k\neq l\leq n_2}
	\widehat V_j^{(k)}\widehat V_j^{(l)}K_{j,kl}.$$
	Because the residual is scalar, the inner product in the KCMD criterion reduces to ordinary multiplication. The nuisance estimator $\widehat f_{n_1,j}$ may be obtained using a deep neural network, random forest, or penalized least squares; the numerical experiments consider neural network and LASSO implementations. The independent unlabeled sample separates nuisance fitting from the labeled observations used in the pairwise statistic, so the construction does not use the additional $\mathcal{U}$-centering adopted in related procedures \citep{lee2020testing,lai2021kernel}.
	
	Conditional on $\mathcal D_1$, $T_{(n_1,n_2),j}$ is a U-statistic of order two. Its oracle version replaces $\widehat f_{n_1,j}$ with $f_j$. Under $H_{0j}$, the first-order projection of the oracle kernel vanishes, while the difference between the estimated and oracle statistics is governed by the nuisance rate and sample size conditions below. This degeneracy motivates the scaling
	$$W_{(n_1,n_2),j}=n_2T_{(n_1,n_2),j}.$$
	
	\subsection{Asymptotic null distribution}
	\label{sec:primary-null-distribution}
	We use the following condition on nuisance estimation.
	\begin{assumption}[Nuisance estimation]\label{assumption:rate}
		For each fixed $j$, let
		\[
		\rho_{n_1,j}^2
		=
		E\!\left[
		\{\widehat f_{n_1,j}(X_{-j})-f_j(X_{-j})\}^2
		\mid\mathcal D_1
		\right],
		\qquad
		f_j(X_{-j})=E(X_j\mid X_{-j}).
		\]
		For a deterministic sequence $a_{n_1,j}\downarrow0$, assume
		\[
		\textnormal{(i)}\quad \rho_{n_1,j}=O_P(a_{n_1,j}),
		\qquad
		\textnormal{(ii)}\quad \sqrt{n_2}\,a_{n_1,j}\to0.
		\]
	\end{assumption}
	
	Assumption~\ref{assumption:rate}(i) gives nuisance consistency, while
	part (ii) makes its error negligible at the null scale. If
	$f_j(x_{-j})=x_{-j}^T\theta_j$ with $s_j=\|\theta_j\|_0$, the lasso gives
	$a_{n_1,j}^2\lesssim s_j\log(p)/n_1$ under standard
	conditions \citep{BickelRitovTsybakov2009}. If $f_j$ is $\beta$-H\"older
	smooth and depends on $d$ coordinates of $x_{-j}$ over a compact domain,
	a ReLU-network estimator can attain
	$a_{n_1,j}^2\lesssim n_1^{-2\beta/(2\beta+d)}$ up to logarithmic factors
	\citep{SchmidtHieber2020}. 
	
	
	\begin{theorem}\label{thm:primary-null-limit}
		Fix $j\in\{1,\ldots,p\}$. Suppose that $H_{0j}$ holds,
		$E|X_j|<\infty$, $E(V_j^4)<\infty$, and
		Assumptions~\ref{assumption:negative},
		\ref{assump:bounded-kernel}, and
		\ref{assumption:rate} hold.
		Let $Z_j=(V_j,\Gamma_j)$ and define
		\[
		J(z,z')
		=
		vv'K(\gamma,\gamma'),
		\qquad
		z=(v,\gamma),\quad z'=(v',\gamma').
		\]
		Define the integral operator $\mathcal T_j$ on $L_2(P_{Z_j})$ by
		\[
		(\mathcal T_j\psi)(z)
		=
		\int J(z,z')\psi(z')\,dP_{Z_j}(z'),
		\]
		and let $\{\zeta_{i,j}\}_{i=1}^{\infty}$ denote its eigenvalues.
		Then, as $(n_1,n_2)\to\infty$,
		\[
		W_{(n_1,n_2),j}
		\Rightarrow
		\sum_{i=1}^{\infty}
		\zeta_{i,j}(G_{i,j}^2-1),
		\]
		where $\{G_{i,j}\}_{i=1}^{\infty}$ are i.i.d. standard Gaussian
		random variables.
	\end{theorem}
	
	Theorem~\ref{thm:primary-null-limit} shows that the null limit is a weighted sum of centered chi-square random variables. The weights $\{\zeta_{i,j}\}_{i=1}^{\infty}$ depend on the unknown distribution $P_{Z_j}$ through the integral operator $\mathcal T_j$, so the limiting distribution is nonpivotal. Rather than estimating the full spectrum, we use the wild bootstrap to approximate its quantiles.
	
	\subsection{Wild bootstrap calibration}
	\label{sec:wild-bootstrap-calibration}
	Because the null limit in Theorem~\ref{thm:primary-null-limit} is nonpivotal, we calibrate the test using a wild bootstrap following \citet{dehling1994random}. Conditional on the observed data, the bootstrap reweights the pairwise summands using multiplier products. Algorithm~\ref{alg:primary-test} summarizes the resulting coordinatewise test. At significance level $\alpha$, we reject $H_{0j}$ if $p_{(n_1,n_2),j,B}\leq\alpha$; otherwise, we do not reject it. The multiplier construction parallels that used for degenerate conditional mean dependence statistics by \citet{lee2020testing}. The resulting bootstrap p-values are also used to construct the e-values in Section~\ref{sec:FDR}.
	
	\begin{algorithm}[tb!]
		\SetAlgoLined
		\DontPrintSemicolon
		\SetKwInOut{Input}{Input}
		\SetKwInOut{Output}{Output}
		\SetKwProg{Proc}{Procedure}{}{}
		
		\caption{Wild bootstrap test}
		\label{alg:primary-test}
		\small
		\Input{
			Coordinate index $j$,
			unlabeled data $\mathcal{D}_1$,
			labeled data $\mathcal{D}_2$,
			prespecified kernel function $K(\cdot,\cdot)$,
			number of bootstrap replications $B$.
		}
		\Output{Bootstrap p-value $p_{(n_1,n_2),j,B}.$}
		
		\BlankLine
		Fit $\widehat f_{n_1,j}$ using $\mathcal D_1$, construct $\widehat V_j^{(k)}$ and $K_{j,kl}$, and compute $W_{(n_1,n_2),j}=n_2T_{(n_1,n_2),j}$ as in Section~\ref{sec:semi-supervised-estimation}.\;
		\For{$b\leftarrow 1$ \KwTo $B$}
		{Generate i.i.d. multipliers $\{\xi_{kj}^{(b)}:1\leq k\leq n_2\}$ with zero mean and unit variance.\;
			Set $W_{(n_1,n_2),j}^{(b)}=n_2T_{(n_1,n_2),j}^{(b)}$, where
			$$T_{(n_1,n_2),j}^{(b)}
			=\frac{1}{n_2(n_2-1)}
			\sum_{1\leq k\neq l\leq n_2}
			\xi_{kj}^{(b)}\xi_{lj}^{(b)}
			\widehat V_j^{(k)}\widehat V_j^{(l)}K_{j,kl}.$$}
		Calculate
		$$p_{(n_1,n_2),j,B}=\frac{1}{B}\sum_{b=1}^{B}I(W_{(n_1,n_2),j}\leq W_{(n_1,n_2),j}^{(b)}),$$
		\Return $p_{(n_1,n_2),j,B}.$\;
	\end{algorithm}
	
	Following \citet{lee2020testing}, we use the following notion of conditional bootstrap weak convergence.
	
	\begin{definition}
		\label{def:bootstrap-convergence}
		Let $Q_n^*$ be a bootstrap random variable and $Q$ a random variable.
		Denote the conditional law of $Q_n^*$ given the observed data by
		$\mathcal L^*(Q_n^*)$ and the law of $Q$ by $\mathcal L(Q)$. We write
		$
		Q_n^*\Rightarrow_P^*Q
		$
		if
		$$
		d_{\mathrm{BL}}
		\left(
		\mathcal L^*(Q_n^*),\mathcal L(Q)
		\right)
		\xrightarrow{P}0,
		$$
		where $d_{\mathrm{BL}}$ is the bounded-Lipschitz metric and convergence
		in probability is under the sampling law of the original data.
	\end{definition}
	
	\begin{assumption}[Bootstrap multipliers]\label{assump:bootstrap-multipliers}
		The multipliers are i.i.d., independent of the data, and satisfy
		$
		E(\xi)=0, E(\xi^2)=1, E(\xi^4)<\infty.
		$
	\end{assumption}
	
	Let $W_{(n_1,n_2),j}^*$ denote a generic draw from the conditional
	multiplier distribution generated by Algorithm~\ref{alg:primary-test}.
	The following theorem shows that its conditional distribution converges
	to the null limit in Theorem~\ref{thm:primary-null-limit}.
	\begin{theorem}\label{thm:bootstrap-consistency}
		Suppose that the conditions of Theorem~\ref{thm:primary-null-limit} and
		Assumption~\ref{assump:bootstrap-multipliers} hold. Then, as
		$(n_1,n_2)\to\infty$,
		\[
		W_{(n_1,n_2),j}^*
		\Rightarrow_P^*
		\sum_{i=1}^{\infty}\zeta_{i,j}(G_{i,j}^2-1),
		\]
		where $\{G_{i,j}\}_{i=1}^{\infty}$ and
		$\{\zeta_{i,j}\}_{i=1}^{\infty}$ are defined in
		Theorem~\ref{thm:primary-null-limit}.
	\end{theorem}
	
	Theorem~\ref{thm:bootstrap-consistency} implies that bootstrap quantiles
	consistently approximate the null critical values. The following corollary
	gives the resulting asymptotic size.
	
	\begin{corollary}\label{cor:asymptotic-size}
		Suppose that the conditions of Theorem~\ref{thm:bootstrap-consistency}
		hold and the limiting null distribution is continuous.
		For every $\alpha\in(0,1)$, as $(n_1,n_2,B)\to\infty$,
		\[
		P\{p_{(n_1,n_2),j,B}\leq\alpha\}\longrightarrow\alpha.
		\]
	\end{corollary}
	Thus, the coordinatewise test is asymptotically exact at level $\alpha$. It supplies the coordinatewise calibration needed before the tests are combined across predictors.
	
	\subsection{Power analysis under fixed and local alternatives}
	\label{sec:primary-power}
	This subsection studies power against the auxiliary alternative
	$H_{1j}':\Lambda_j>0$, rather than against all departures from the scientific
	null $H_{0j}$. Since
	$\Lambda_j=\|E\{V_j\Phi(\Gamma_j)\}\|_{\mathcal H_K}^2$, where
	$\Phi(\gamma)=K(\gamma,\cdot)\in\mathcal H_K$, the results apply to
	alternatives detected by the identity specification. They do not cover
	violations of conditional independence for which $\Lambda_j=0$.
	
	\begin{theorem}
		\label{thm:fixed-power}
		Fix $j\in\{1,\ldots,p\}$. Suppose that $H_{1j}':\Lambda_j>0$ holds,
		$E|X_j|<\infty$, $E(V_j^4)<\infty$, and
		Assumptions~\ref{assumption:negative}, \ref{assump:bounded-kernel},
		\ref{assumption:rate}(i), and
		\ref{assump:bootstrap-multipliers} hold. Then, for any integer sequence
		$B\geq1$, as $(n_1,n_2)\to\infty$,
		$
		p_{(n_1,n_2),j,B}\xrightarrow{P}0
		$.
		Consequently, for every fixed $\alpha\in(0,1)$,
		\[
		P\{p_{(n_1,n_2),j,B}\leq\alpha\}\longrightarrow1.
		\]
	\end{theorem}
	
	We next describe alternatives that approach the auxiliary null at a
	controlled rate. Let $\{P_{n_2}\}$ be a sequence of joint distributions
	and let $E_{n_2}$ denote expectation under $P_{n_2}$. Throughout this
	subsection, $X_j$, $X_{-j}$, $Y$, and $\Gamma_j$ are understood under
	$P_{n_2}$, and the sequence index is suppressed on these basic random
	variables. Define
	\[
	V_j^{[n_2]}
	=
	X_j-E_{n_2}(X_j\mid X_{-j}),
	\qquad
	\mu_j^{[n_2]}
	=
	E_{n_2}\{V_j^{[n_2]}\Phi(\Gamma_j)\}.
	\]
	Here $\Gamma_j=(\phi(Y),X_{-j})$ is defined as before, although its
	distribution may vary with $P_{n_2}$.
	The corresponding population criterion satisfies
	$\Lambda_j^{[n_2]}=\|\mu_j^{[n_2]}\|_{\mathcal H_K}^2$.
	
	\begin{assumption}[Local mean-element drift]
		\label{ass:local-primary}
		For a sequence $\varrho_{n_2}\downarrow0$, suppose that
		\[
		\mu_j^{[n_2]}
		=
		\varrho_{n_2}h_j+o(\varrho_{n_2})
		\quad\text{in }\mathcal H_K,
		\qquad
		0<\|h_j\|_{\mathcal H_K}<\infty.
		\]
		Let
		$A_j^{0,[n_2]}=V_j^{[n_2]}\Phi(\Gamma_j)-\mu_j^{[n_2]}$ and let
		$\Sigma_j^{[n_2]}$ be its covariance operator on $\mathcal H_K$.
		Assume that $\Sigma_j^{[n_2]}$ converges in trace norm to a trace-class
		operator $\Sigma_j$, that
		\[
		\sup_{n_2}
		E_{n_2}\|A_j^{0,[n_2]}\|_{\mathcal H_K}^4<\infty,
		\]
		and that Assumption~\ref{assumption:rate} holds along
		$\{P_{n_2}\}$, with $f_j(X_{-j})$ replaced by
		$E_{n_2}(X_j\mid X_{-j})$.
	\end{assumption}
	
	Assumption~\ref{ass:local-primary} specifies the departure directly in
	the mean element measured by the statistic. To display the perturbation
	explicitly, write $X_j^{[n_2]}$ for $X_j$ under $P_{n_2}$. Under a fixed
	joint distribution of $(Y,X_{-j},\varepsilon_j,\Delta_j)$, with expectation
	denoted by $E_0$, one concrete construction is
	\[
	X_j^{[n_2]}=f_j(X_{-j})+\varepsilon_j+
	\varrho_{n_2}\Delta_j(Y,X_{-j}),
	\]
	where
	$
	E_0(\varepsilon_j\mid Y,X_{-j})=0,
	E_0\{\Delta_j(Y,X_{-j})\mid X_{-j}\}=0.
	$
	Then
	\[
	E_{n_2}(X_j^{[n_2]}\mid X_{-j})=f_j(X_{-j}),
	\qquad
	E_{n_2}(V_j^{[n_2]}\mid Y,X_{-j})
	=
	\varrho_{n_2}\Delta_j(Y,X_{-j}),
	\]
	and hence
	\[
	\mu_j^{[n_2]}
	=
	\varrho_{n_2}
	E_0\{\Delta_j(Y,X_{-j})\Phi(\Gamma_j)\}.
	\]
	If $E_0(\varepsilon_j^4)<\infty$, $E_0(\Delta_j^4)<\infty$, and
	\[
	0<
	\left\|
	E_0\{\Delta_j(Y,X_{-j})\Phi(\Gamma_j)\}
	\right\|_{\mathcal H_K}
	<\infty,
	\]
	then, since $K$ is bounded and $\varrho_{n_2}\to0$, the required uniform
	fourth-moment bound and trace-norm covariance convergence hold.
	
	\begin{theorem}
		\label{thm:local-power}
		Fix $j\in\{1,\ldots,p\}$ and $\alpha\in(0,1)$, and suppose that
		Assumptions~\ref{assumption:negative},
		\ref{assump:bounded-kernel}, \ref{assump:bootstrap-multipliers}, and
		\ref{ass:local-primary} hold along $\{P_{n_2}\}$. Let
		$
		\mathbb G_j\sim N_{\mathcal H_K}(0,\Sigma_j),
		$
		and define
		\[
		Q_{0,j}
		=
		\|\mathbb G_j\|_{\mathcal H_K}^2-\operatorname{tr}(\Sigma_j),
		\qquad
		Q_{\tau,j}
		=
		\|\mathbb G_j+\tau h_j\|_{\mathcal H_K}^2-\operatorname{tr}(\Sigma_j),
		\quad \tau\geq0.
		\]
		Assume that $q_{j,1-\alpha}$ is the unique $(1-\alpha)$-quantile
		of $Q_{0,j}$ and that the distribution function of $Q_{0,j}$ is
		continuous at $q_{j,1-\alpha}$.
		Then, as $(n_1,n_2,B)\to\infty$:
		\begin{enumerate}
			\item If $\sqrt{n_2}\varrho_{n_2}\to\infty$, then
			\[
			P\{p_{(n_1,n_2),j,B}\leq\alpha\}\longrightarrow1.
			\]
			\item If $\sqrt{n_2}\varrho_{n_2}\to\tau\in(0,\infty)$, then
			\[
			P\{p_{(n_1,n_2),j,B}\leq\alpha\}
			\longrightarrow
			P(Q_{\tau,j}>q_{j,1-\alpha}).
			\]
			\item If $\sqrt{n_2}\varrho_{n_2}\to0$, then
			\[
			P\{p_{(n_1,n_2),j,B}\leq\alpha\}\longrightarrow\alpha.
			\]
		\end{enumerate}
	\end{theorem}
	
	If the singleton family $\{g(x)=x\}$ is coordinate exhaustive, these
	results apply directly to the scientific marginal coordinate hypothesis.
	Within the identity-detectable class, the detection boundary is
	$n_2^{-1/2}$ for $\|\mu_j^{[n_2]}\|_{\mathcal H_K}$, or equivalently
	$n_2^{-1}$ for $\Lambda_j^{[n_2]}$. Coordinate exhaustiveness alone does
	not ensure that every local sequence admits the mean-element drift in
	Assumption~\ref{ass:local-primary}.

	\section{Simultaneous inference via e-values}
	\label{sec:FDR}
	Section~\ref{sec:Test} constructs an asymptotically valid bootstrap p-value for each marginal coordinate hypothesis. To identify active predictors, the \(p\) coordinate hypotheses must be tested simultaneously. The resulting p-values can be strongly dependent because the coordinatewise statistics use the same labeled sample and overlapping predictor sets. The e-BH procedure of \citet{wang2022false} controls the FDR under arbitrary dependence for valid e-values. We therefore calibrate the bootstrap p-values into truncated e-values and apply e-BH to the resulting collection.
	
	\subsection{Calibration and the e-BH procedure}
	\label{sec:evalue-calibration}
	We begin with the terminology needed for calibration. Following the definitions of \citet{vovk2021values} and adopting the convention of \citet{wang2022false}, we use p-value and e-value for both random variables and their realized values. Under a null hypothesis $H_0$, a valid p-value is a random variable $\mathbb{P}$ taking values in $[0,1]$ and satisfying
	$P_{H_0}(\mathbb{P}\leq\alpha)\leq\alpha$
	for all $\alpha\in(0,1)$, whereas a valid e-value is an extended nonnegative random variable $\mathbb{E}$ satisfying
	$$E_{H_0}(\mathbb{E})\leqslant 1.$$
	Thus, evidence against a null is represented by a small p-value or a large e-value. Proposition 2.1 of \citet{vovk2021values} shows that $h(\mathbb P)$ is a valid e-value whenever $\mathbb P$ is a valid p-value and $h$ is a calibrator: a decreasing function $h:[0,1]\rightarrow[0,\infty]$ satisfying $\int_0^1 h(x)\,dx\leq 1$.
	Examples include $h(x)=\upsilon x^{\upsilon-1}$ for $\upsilon\in(0,1)$ and the integrated calibrator $h(x)=\frac{1-x+x\log x}{x(\log x)^2}$. Further discussion is given by \citet{vovk2021values}.

	\begin{algorithm}[tbh!]
		\SetAlgoLined
		\DontPrintSemicolon
		\SetKwInOut{Input}{Input}
		\SetKwInOut{Output}{Output}
		\SetKwProg{Proc}{Procedure}{}{}
		\SetKwComment{Phase}{}{}
		
		\caption{Simultaneous inference via e-values}
		\label{alg:fdr-control}
		\Input{
			bootstrap p-values $\{p_{(n_1,n_2),j,B}\}_{j=1}^p$,
			FDR level $\alpha$, continuous admissible calibrator $h$.}
		\Output{Rejection set $\mathcal{E}(e)$.}
		\BlankLine
		\Phase{\textnormal{\textbf{Calibration}}}
		\Indp
		Define the truncated calibrator\hfill\break
		\makebox[\dimexpr\linewidth-3\algomargin-\algoskipindent\relax][c]{$\displaystyle
			h_{trun}(x)=\min\{h(x),p/\alpha\},\qquad 0\leq x\leq1.
			$}\;
		\For{$j\leftarrow 1$ \KwTo $p$}
		{Set $e_{(n_1,n_2),j,B}=h_{trun}(p_{(n_1,n_2),j,B})$.\;
		}
		\Indm
		\BlankLine
		\Phase{\textnormal{\textbf{e-BH}}}
		\Indp
		Order the e-values as $e_{(n_1,n_2),[1],B}\geq\cdots\geq e_{(n_1,n_2),[p],B}$.\;
		With $\mathcal K=\{1,\ldots,p\}$, compute\hfill\break
		\makebox[\dimexpr\linewidth-3\algomargin-\algoskipindent\relax][c]{$\displaystyle
			k_e^{*}=\max \{ k\in\mathcal{K} : \frac{ke_{(n_1,n_2),[k],B}}{p}\geq \frac{1}{\alpha }\} .
			$}\;
		If the set is empty, set $k_e^*=0$ and $\mathcal E(e)=\varnothing$.
		Otherwise, reject the hypotheses satisfying
		$e_{(n_1,n_2),j,B}\ge p/(\alpha k_e^*)$ and denote their index set by
		$\mathcal E(e)$.\;
		\Return $\mathcal{E}(e)$.\;
		\Indm
	\end{algorithm}

	\subsection{Asymptotic FDR guarantee}
	\label{sec:asymptotic-fdr-guarantee}
	Corollary~\ref{cor:asymptotic-size} establishes the asymptotic validity of the coordinatewise bootstrap p-values. The following definition formalizes the corresponding validity notion for e-values.
	\begin{definition}\label{def:asymptotic-e-value}
		A sequence of extended nonnegative random variables $\{\mathbb{E}_n\}$ is an asymptotically valid e-value sequence under the null hypothesis $H_0$ if
		$$\limsup_{n\rightarrow{} \infty} E_{H_0}(\mathbb{E}_n)\leq 1.$$
	\end{definition}
	
	Because the bootstrap p-values are only asymptotically valid, we truncate the calibrator to obtain the expectation bound underlying the asymptotic validity of the resulting e-values. Algorithm~\ref{alg:fdr-control} summarizes the resulting procedure including e-BH. 
	Truncation is needed for the expectation argument but does not alter the e-BH rejection set. The following proposition makes it explicit.
	\begin{proposition}\label{prop:calibrator-rejection-equivalence}
		Let $e'_j=h(p_j)$ and $e_j=h_{trun}(p_j)$, and let
		$\mathcal E(e')$ and $\mathcal E(e)$ be the rejection sets
		obtained from the respective e-values by the e-BH step-up rule in
		Algorithm~\ref{alg:fdr-control} at level $\alpha$. Then
		$
		\mathcal E(e')=\mathcal E(e).
		$
	\end{proposition}

	Let $\mathcal{E}$ denote the simultaneous testing procedure, and let $\mathcal{R}_{\mathcal{E}}$ and $\mathcal{F}_{\mathcal{E}}$ denote its numbers of rejections and false rejections, respectively. The following theorem first establishes the asymptotic validity of the calibrated e-values and then applies e-BH to obtain FDR control.
	\begin{theorem}\label{thm:asymptotic-fdr}
		Suppose that the conditions of Corollary~\ref{cor:asymptotic-size}
		hold for every true null coordinate. Then, for each such coordinate,
		\[
		\limsup_{(n_1,n_2,B)\to\infty}
		E\{e_{(n_1,n_2),j,B}\}\leq1.
		\]
		Moreover,
		\[
		\limsup_{(n_1,n_2,B)\to\infty}FDR
		=
		\limsup_{(n_1,n_2,B)\to\infty}
		E\!\left\{
		\frac{\mathcal F_{\mathcal E}}
		{\mathcal R_{\mathcal E}\vee1}
		\right\}
		\leq\alpha.
		\]
	\end{theorem}
	
	Theorem~\ref{thm:asymptotic-fdr} therefore controls the limiting FDR
	without imposing a particular dependence structure across coordinates.
	A corresponding guarantee for level-specific boosted e-values follows by combining our proof with Theorem~3 of \citet{wang2022false}; details are omitted because this case is not our primary focus.
	
	The same simultaneous inference construction applies to the multiple transformation extension in Section~\ref{sec:exhaustive-extension}. Its bootstrap p-values, defined in Supplementary Material~\ref*{app:exhaustive-test}, can be transformed by the same truncated calibrator and combined through e-BH. The corresponding asymptotic results are given in Supplementary Material~\ref*{app:exhaustive-simultaneous}.

	\section{Numerical experiments}
	\label{sec:Simulation}
	We evaluate the finite-sample performance of FMCT for coordinatewise and simultaneous testing with Euclidean and metric space responses. The first two subsections report coordinatewise results for linear and nonlinear regression models and for distributional, SPD matrix, and spherical responses; the final subsection reports empirical FDR and power for simultaneous testing.
	
	FMCT-Lasso estimates each $f_j(X_{-j})$ by LASSO with its tuning parameter selected by 10-fold cross-validation. FMCT-NN uses a multilayer perceptron with depth $L=3$ and width $W=50$. All reported experiments use the bounded Gaussian kernel
	$K(\gamma,\gamma')=\exp\{-d(\gamma,\gamma')^2/\sigma\}$, where $\gamma,\gamma' \in \mathcal{H} \times \mathbb{R}^{p-1}$.
	Following \citet{lai2021kernel}, the bandwidth $\sigma$ is set to the sample median of the pairwise squared distances.
	In the product space $\mathcal{H}\times\mathbb{R}^{p-1}$, the covariate contribution to the distance between two observations can dominate the response contribution when $p$ is moderately large:
	$$d_{\mathcal{H} \times \mathbb{R}^{p-1}}^2(\Gamma_j^{(k)},\Gamma_j^{(l)})
	=d_{\mathcal Y}(Y^{(k)},Y^{(l)})
	+\|X_{-j}^{(k)}-X_{-j}^{(l)}\|_2^2.$$
	We balance the two components by multiplying the covariate distance by the dimension-based factor
	$$\operatorname{dim}(\mathcal{Y} )/\operatorname{dim}(\mathbb{R}^{p-1}) \approx \operatorname{dim}(\mathcal{Y} )/p,$$
	where $\operatorname{dim}(\mathcal{Y})$ denotes the effective response dimension used in computation and is specified for each response space below. Throughout the coordinatewise experiments, $n_1:n_2=4:1$. Each coordinatewise experiment uses 500 Monte Carlo replications, $B=500$ bootstrap replications, independent $N(0,1)$ multipliers, and nominal level $\alpha=0.05$.

	\subsection{Euclidean responses}
	\label{sec:euclidean-responses}
	We first consider two scalar-response settings, so $\mathcal{Y}=\mathbb R$ and $\operatorname{dim}(\mathcal{Y})=1$. We compare FMCT with the decorrelated score procedure $T^{NL}$-lasso \citep{ning2017general} and the model-free procedure $W_n$-Lasso of \citet{guo2024model}.
	\begin{example}
		Let $X=(Z^\top,W^\top)^\top\sim N_{20}(0,\Sigma)$, where $Z\in\mathbb R^4$, $W\in\mathbb R^{16}$, and $\Sigma_{rs}=0.3^{|r-s|}$ for $r,s=1,\ldots,20$. The response is generated from
		$$Y=\beta_Z^\top Z+\beta_W^\top W+\epsilon,$$
		where $\epsilon\sim N(0,0.5^2)$ is independent of $X$. Every component of $\beta_Z$ equals $1/\sqrt{2}$, whereas every component of $\beta_W$ equals zero. Thus, $X_1,\ldots,X_4$ are active, whereas the remaining coordinates are inactive.
	\end{example}
	The second example is adapted from \citet{guo2024model}.
	\begin{example}
		Consider
		$$Y=\frac{X_1+X_2}{0.5+(1.5+X_3+X_4)^2}+0.1\epsilon,$$
		where $X\sim N_{20}(0,\Sigma)$ with $\Sigma_{rs}=0.01^{|r-s|}$ for $r,s=1,\ldots,20$, and $\epsilon\sim N(0,1)$ is independent of $X$.
		Thus, $X_1,X_2,X_3$, and $X_4$ are active.
	\end{example}

	Table~\ref{table:Example 1,2} reports coordinatewise rejection frequencies. In Example 1, FMCT-NN and FMCT-Lasso attain power above $0.9$ for all four active coordinates even at $n_2=100$, while their null-coordinate rejection frequencies are broadly comparable to those of the two competing methods.
	In Example 2, both FMCT-NN and FMCT-Lasso detect the nonlinear contributions of $X_3$ and $X_4$, with power increasing from approximately $0.92$ at $n_2=100$ to $1$ at $n_2=200$. Their average power over these two coordinates is slightly lower than that of $W_n$-Lasso at $n_2=100$ but becomes comparable at the larger sample sizes.
	Their null-coordinate rejection frequencies are generally close to the nominal level, with occasional mild overrejection. By contrast, $T^{NL}$-lasso rejects $X_3$ and $X_4$ at frequencies close to the nominal level, indicating little power against these nonlinear effects.
	
	{\spacingset{1}
		\begin{table}[!htbp]
			\renewcommand{\arraystretch}{1.2}
			\centering
			\begin{threeparttable}
				\caption{Coordinatewise rejection frequencies in Examples 1 and 2 with $p=20$ at the nominal level $\alpha=0.05$. Underlined coordinates are active.}
				\begin{tabular}{c c c c c c c c c c}
					\hline
					\multicolumn{10}{c}{Example 1: active coordinates $X_1,X_2,X_3$, and $X_4$} \\
					\hline
					$n_2$& Method & $\underline{X_1}$ & $\underline{X_2}$ & $\underline{X_3}$ & $\underline{X_4}$ & $X_5$ & $X_{10}$ & $X_{15}$ & $X_{20}$ \\
					\hline
					\multirow{4}{*}{100} & FMCT-NN & 0.944 & 0.914 & 0.922 & 0.916 & 0.060 & 0.042 & 0.058 & 0.046\\
					& FMCT-Lasso & 0.946 & 0.920 & 0.908 & 0.932 & 0.082 & 0.048 & 0.048 & 0.054\\
					&$W_n$-Lasso& 1.000& 1.000 & 1.000 &1.000 & 0.058 & 0.060 & 0.052  & 0.054\\
					&$T^{NL}$-lasso& 1.000 & 1.000 & 1.000 & 1.000 & 0.072 & 0.048 & 0.062 &0.086 \\
					\hline
					\multirow{4}{*}{150} & FMCT-NN & 0.994 & 0.982 & 0.996 & 0.996 & 0.044 & 0.044 & 0.044 & 0.072\\
					& FMCT-Lasso & 0.996 & 0.984 & 0.980 & 0.990 & 0.078 & 0.050 & 0.068 & 0.054\\
					& $W_n$-Lasso & 1.000 & 1.000 & 1.000 & 1.000 & 0.066 & 0.052 & 0.054 &0.038\\
					& $T^{NL}$-lasso & 1.000 & 1.000 & 1.000 & 1.000 & 0.082 & 0.062 & 0.058 & 0.050\\
					\hline
					\multirow{4}{*}{200} & FMCT-NN & 1.000 & 0.996 & 0.998 & 0.996 & 0.070 & 0.044 & 0.068 & 0.046 \\
					& FMCT-Lasso & 1.000 & 1.000 & 0.996 & 1.000 & 0.058 & 0.062 & 0.072 & 0.058\\
					& $W_n$-Lasso & 1.000 & 1.000 & 1.000 & 1.000 & 0.060& 0.036 &0.034 &0.042\\
					& $T^{NL}$-lasso & 1.000 & 1.000 & 1.000 & 1.000 & 0.046 & 0.050 & 0.068 &0.070  \\
					\hline
					\multicolumn{10}{c}{Example 2: active coordinates $X_1,X_2,X_3$, and $X_4$} \\
					\hline
					$n_2$& Method & $\underline{X_1}$ & $\underline{X_2}$ & $\underline{X_3}$ & $\underline{X_4}$ & $X_{5}$ & $X_{10}$ & $X_{15}$ & $X_{20}$ \\
					\hline
					\multirow{4}{*}{100} & FMCT-NN & 1.000 & 1.000 & 0.920 & 0.916 & 0.048 & 0.046 & 0.048 & 0.050\\
					& FMCT-Lasso & 1.000 & 1.000 & 0.912 & 0.918 & 0.054 & 0.050 & 0.054 &0.056\\
					&$W_n$-Lasso& 1.000& 1.000 & 0.932 & 0.914 & 0.074 & 0.064 & 0.048 & 0.050\\
					&$T^{NL}$-lasso& 1.000 & 1.000 & 0.058 &0.068 & 0.080 & 0.058 & 0.078 & 0.060\\
					\hline
					\multirow{4}{*}{150} & FMCT-NN & 1.000 & 1.000 & 0.996 & 0.990 & 0.060 & 0.052 & 0.070 & 0.062\\
					& FMCT-Lasso & 1.000 & 1.000 & 0.996 & 0.990 & 0.064 & 0.052 & 0.062 &0.034\\
					& $W_n$-Lasso & 1.000 & 1.000 & 0.990 & 0.996 & 0.044 & 0.038 &0.048 &0.024\\
					& $T^{NL}$-lasso & 1.000 & 1.000 & 0.064 & 0.076 & 0.054 &0.042 & 0.062 & 0.044 \\
					\hline
					\multirow{4}{*}{200} & FMCT-NN & 1.000 & 1.000 & 1.000 & 1.000 & 0.060 & 0.072 & 0.044 & 0.046\\
					& FMCT-Lasso & 1.000 & 1.000 & 1.000 & 1.000 & 0.056 & 0.052 & 0.064 & 0.054\\
					& $W_n$-Lasso & 1.000 & 1.000 & 1.000 & 1.000 & 0.036 & 0.040 & 0.038& 0.044\\
					& $T^{NL}$-lasso & 1.000 & 1.000 & 0.060 & 0.050 & 0.066 & 0.040 & 0.040 &0.064 \\
					\hline
				\end{tabular}
				\label{table:Example 1,2}
			\end{threeparttable}
		\end{table}
	}

	\subsection{Metric space responses}
	\label{sec:metric-space-responses}
	We next examine three kinds of non-Euclidean responses used in Fr\'{e}chet regression and sufficient dimension reduction \citep{qiu2024random,ying2022frechet}: probability distributions, SPD matrices, and spherical data. 
	
	\subsubsection{Distributional responses}
	\label{sec:distribution-valued-responses}
	Let $\mathcal{P}$ denote the space of probability distributions on $\mathbb{R}$ with finite second moments. For $Y_1,Y_2\in\mathcal{P}$, their squared Wasserstein distance is
	$$d_W^{2}(Y_1,Y_2)=\int_{0}^{1} \{Q_{Y_1}(t)-Q_{Y_2}(t)\}^{2} \,dt,$$
	where $Q_{Y_1}$ and $Q_{Y_2}$ denote the corresponding quantile functions. Since $(\mathcal{P},d_W)$ is infinite-dimensional, we evaluate the quantile functions on a finite grid, take the number of grid points as $\operatorname{dim}(\mathcal{Y})$, and approximate $d_W$ by the Euclidean distance between the resulting quantile vectors.
	
	\begin{example}
		Consider
		$$Q_Y(t)=\mu_Y+\sigma_Y\Phi^{-1}(t),\qquad t\in(0,1),$$
		where $Q_Y$ is the quantile function of $Y$ and $\Phi^{-1}$ is the standard normal quantile function. 
		Conditional on $X$, $\mu_Y\sim N\{D_1(X),0.1^2\}$ and $\sigma_Y=|D_2(X)|$, where $
		D_1(X)=(X_1^2+X_2^2)^{1/2}\log\{(X_1^2+X_2^2)^{1/2}\},
		D_2(X)=\sin\{0.5\pi(X_4+X_5)\}+X_4^2
		$. We generate $X\sim\operatorname{Unif}([0,1]^{20})$. Thus, $X_1,X_2,X_4$, and $X_5$ are active.
	\end{example}
	\subsubsection{SPD matrix responses}
	\label{sec:spd-matrix-responses}
	Let $S_{+}^m$ be the set of $m\times m$ SPD matrices. For any $Y\in S_{+}^m$, let $P$ denote the lower-triangular Cholesky factor satisfying $PP^\top=Y$. For $Y_1,Y_2\in S_{+}^m$, the Log-Cholesky distance is
	$$d_L(Y_1,Y_2)=[\|\lfloor P_1\rfloor - \lfloor P_2\rfloor \Vert_F^2 +\|\log \mathbb{D} (P_1) - \log\mathbb{D} (P_2)\Vert_F^2 ]^{1/2},$$
	where $\mathbb{D}(A)$ and $\lfloor A\rfloor$ denote the diagonal and strictly lower-triangular parts of a matrix $A$, respectively, $\log(\cdot)$ is the matrix logarithm, and $\|\cdot\Vert_F$ is the Frobenius norm. Thus, $(S_{+}^m, d_L)$ is a metric space with $\operatorname{dim}(S_{+}^m)=m(m+1)/2$.

	For data generation, write $A\sim N_{m\times m}^{\mathrm{sym}}(M,\sigma^2)$ if $A=\sigma C+M$, where $M$ is an $m\times m$ symmetric matrix, the variables $C_{rr}\sim N(0,1)$ and $C_{rs}\sim N(0,1/2)$, $r<s$, are mutually independent, and $C_{sr}=C_{rs}$.
	\begin{example}
		Let
		$M(X)=\begin{pmatrix}
			1 & \rho(X) \\
			\rho(X) & 1
		\end{pmatrix}, \rho(X)=3\sin X_1+3\sin X_2,$
		and generate
		$$\log(Y)\sim N_{2\times2}^{\mathrm{sym}}\{M(X),0.2^2\}.$$
		We first generate $Z\sim N_{20}(0,\Sigma)$ with $\Sigma_{rs}=0.9^{|r-s|}$ for $r,s=1,\ldots,20$. Then, let $X_j=2\Phi(Z_j)$ for $j=1,\ldots,20$, where $\Phi$ is the standard normal distribution function. $X_1$ and $X_2$ are active.
		
	\end{example}
	\subsubsection{Spherical responses}
	\label{sec:spherical-responses}
	Let $(\mathbb{S}^2,d_g)$ denote the unit sphere equipped with the geodesic distance
	$d_g(Y_1,Y_2)=\arccos(Y_1^\top Y_2),$
	for $Y_1,Y_2\in\mathbb{S}^2$. Here $\operatorname{dim}(\mathbb{S}^2)=2$.
	\begin{example}
		Consider
		\[
		\begin{aligned}
			Y=\big(&\sin(f_1(X)+\epsilon_1)\sin(f_2(X)+\epsilon_2),\\
			&\sin(f_1(X)+\epsilon_1)\cos(f_2(X)+\epsilon_2),
			\cos(f_1(X)+\epsilon_1)\big)^T,
		\end{aligned}
		\]
		where $f_1(X)=X_1+X_2$ and $f_2(X)=X_{19}+X_{20}$. The noises satisfy $\epsilon_k\stackrel{\text{i.i.d.}}{\sim}N(0,0.2^2)$ for $k=1,2$ and are independent of $X$.
		We first generate $Z\sim N_{20}(0,\Sigma)$ with $\Sigma_{rs}=0.5^{|r-s|}$ for $r,s=1,\ldots,20$. Then, let $X_j=\Phi(Z_j)$ for $j=1,\ldots,20$. $X_1,X_2,X_{19}$, and $X_{20}$ are active.
	\end{example}
	\FloatBarrier
	
	{\spacingset{1}
		\begin{table}[!t]
			\renewcommand{\arraystretch}{1.2}
			\small
			\setlength{\tabcolsep}{5pt}
			\centering
			\begin{threeparttable}
				\caption{Coordinatewise rejection frequencies for metric space responses with $p=20$ at the nominal level $\alpha=0.05$. Underlined coordinates are active.}
				\begin{tabular}{c c c c c c c c c c}
					\hline
					\multicolumn{10}{c}{Example 3: active coordinates $X_1,X_2,X_4$, and $X_5$} \\
					\hline
					& $n_2$ & $\underline{X_1}$ & $\underline{X_2}$  & $X_3$ &$\underline{X_4}$ & $\underline{X_5}$ & $X_6$ & $X_{10}$ & $X_{20}$ \\
					\hline
					& 40 & 0.382 & 0.378 & 0.038 & 0.870 & 0.140 & 0.040 &0.050 &0.058\\
					FMCT-NN & 120 & 0.930 & 0.926 & 0.072 & 1.000 & 0.564 & 0.050 &0.040&0.056\\
					& 200 & 1.000 & 0.998 &0.046 & 1.000 & 0.870 & 0.064 & 0.034 & 0.056\\
					\hline
					& 40 & 0.402 & 0.386 & 0.050 & 0.882 & 0.152 & 0.042 &0.048&0.056\\
					FMCT-Lasso & 120 & 0.918 & 0.910 & 0.032 & 1.000 & 0.530 & 0.040 &0.064&0.046\\
					& 200 & 0.998 & 0.998 &0.046 & 1.000 & 0.890 & 0.040 & 0.044 & 0.046\\
					\hline
					\multicolumn{10}{c}{Example 4: active coordinates $X_1$ and $X_2$} \\
					\hline
					& $n_2$ & $\underline{X_1}$ & $\underline{X_2}$ & $X_3$ & $X_4$ & $X_{10}$ & $X_{11}$ & $X_{19}$ & $X_{20}$ \\
					\hline
					& 40 & 0.372 & 0.264  & 0.072 &  0.068 &0.044 &0.072 & 0.048 & 0.058\\
					FMCT-NN & 120 & 0.794 & 0.636  & 0.064 & 0.072 & 0.042 &0.056 & 0.042 &0.040 \\
					& 200 & 0.948 &  0.844 & 0.062 & 0.056 & 0.052 & 0.052 & 0.040 & 0.042\\
					\hline
					& 40 & 0.380 & 0.300 & 0.076 & 0.066 & 0.078 & 0.060 &0.068&0.064\\
					FMCT-Lasso & 120 & 0.804 & 0.642 & 0.094 & 0.096 & 0.064 & 0.056 &0.046&0.044\\
					& 200 & 0.956 & 0.868 &0.074 & 0.064 & 0.060 & 0.068 & 0.048 & 0.066\\
					\hline
					\multicolumn{10}{c}{Example 5: active coordinates $X_1,X_2,X_{19}$, and $X_{20}$} \\
					\hline
					& $n_2$ & $\underline{X_1}$ & $\underline{X_2}$ & $X_3$ & $X_8$ & $X_{13}$ & $X_{18}$ & $\underline{X_{19}}$ & $\underline{X_{20}}$ \\
					\hline
					& 40 &0.854 &0.762 &0.062 & 0.070 & 0.054 & 0.046 & 0.526 &0.578\\
					FMCT-NN & 120 & 1.000 & 0.998 & 0.064 &  0.048 & 0.046 & 0.072 & 0.978 & 0.996\\
					& 200 & 1.000 & 1.000 & 0.080 & 0.032 &0.050 & 0.054& 1.000 & 1.000\\
					\hline
					& 40 & 0.856 & 0.812 & 0.076 & 0.042 & 0.046 & 0.096 &0.514&0.608\\
					FMCT-Lasso & 120 & 1.000 & 0.998 & 0.058 & 0.048 & 0.048 & 0.064 &0.980&0.992\\
					& 200 & 1.000 & 1.000 &0.072 & 0.052 & 0.062 & 0.050 & 1.000 & 1.000\\
					\hline
				\end{tabular}
				\label{table:Example 3 4 5}
			\end{threeparttable}
		\end{table}
	}

	Table~\ref{table:Example 3 4 5} reports the coordinatewise rejection frequencies. For the distributional response in Example 3, power increases with $n_2$ for every active coordinate, although the contribution of $X_5$ is more difficult to detect. The null-coordinate rejection frequencies remain generally close to the nominal level across the reported sample sizes. Example 4 shows a similar improvement in power for the SPD matrix response; its null-coordinate rejection frequencies are mostly near the nominal level, although FMCT-Lasso exhibits some overrejection at $n_2=120$. In the spherical setting of Example 5, power is high at the two larger sample sizes, while the null-coordinate rejection frequencies are mostly close to the nominal level with occasional mild overrejection. Overall, the results show that FMCT can detect predictor contributions across the three response geometries, with performance improving as more labeled observations become available.

	\subsection{Simultaneous testing}
	\label{sec:simultaneous-testing-performance}
	
	We simultaneously test all $p=20$ coordinate hypotheses using 100 Monte Carlo replications at nominal FDR levels $\alpha=0.1$ and $0.2$. Empirical FDR and power are the average false discovery and true positive proportions, respectively; other settings are as above. We use the truncated, level-specific boosted p-to-e transformation based on $h_{\lambda}(x)=\lambda x^{\lambda-1}$, with $\lambda=1/2$ and boosting factor $b_{\alpha}=\sqrt{2/\alpha}$ \citep{wang2022false}.
	
	For Euclidean responses, we compare FMCT with model-X knockoffs \citep{candes2018panning} and the $W_n$-Lasso procedure of \citet{guo2024model}. We use the default lasso coefficient-difference statistic for knockoffs and the default setting for $W_n$-Lasso. The experiments include the linear model in Example~1 and the following model from \citet{guo2024model}.

	{\spacingset{1}
		\begin{table}[!htbp]
			\caption{Empirical FDR and power for the Euclidean-response models in
				Examples~1 and~6 with $p=20$ at nominal FDR levels $\alpha=0.1$ and $\alpha=0.2$.}
			\centering
			\small
			\setlength{\tabcolsep}{4pt}
			\renewcommand{\arraystretch}{1.15}
			\begin{tabular*}{\textwidth}
				{
					@{\extracolsep{\fill}}
					l
					cccc
					@{\hspace{10pt}}
					cccc
					@{}
				}
				\toprule
				& \multicolumn{4}{c}{Example~1}
				& \multicolumn{4}{c}{Example~6} \\
				\cmidrule(lr){2-5}
				\cmidrule(lr){6-9}
				
				& \multicolumn{2}{c}{$\alpha=0.1$}
				& \multicolumn{2}{c}{$\alpha=0.2$}
				& \multicolumn{2}{c}{$\alpha=0.1$}
				& \multicolumn{2}{c}{$\alpha=0.2$} \\
				\cmidrule(lr){2-3}
				\cmidrule(lr){4-5}
				\cmidrule(lr){6-7}
				\cmidrule(lr){8-9}
				
				Method
				& FDR & Power
				& FDR & Power
				& FDR & Power
				& FDR & Power \\
				\midrule
				
				\multicolumn{9}{c}{$n_2=120$} \\
				\midrule
				FMCT-NN
				& 0.018 & 0.695
				& 0.025 & 0.773
				& 0.018 & 0.950
				& 0.029 & 0.977 \\
				
				FMCT-Lasso
				& 0.018 & 0.853
				& 0.103 & 0.975
				& 0.012 & 0.583
				& 0.028 & 0.683 \\
				
				Knockoff
				& 0.135 & 0.978
				& 0.221 & 0.988
				& 0.197 & 0.623
				& 0.248 & 0.653 \\
				
				$W_n$-Lasso
				& 0.208 & 1.000
				& 0.209 & 1.000
				& 0.210 & 1.000
				& 0.210 & 1.000 \\
				
				\midrule
				\multicolumn{9}{c}{$n_2=160$} \\
				\midrule
				FMCT-NN
				& 0.019 & 0.868
				& 0.030 & 0.935
				& 0.010 & 0.997
				& 0.014 & 1.000 \\
				
				FMCT-Lasso
				& 0.019 & 0.963
				& 0.111 & 0.995
				& 0.003 & 0.767
				& 0.041 & 0.847 \\
				
				Knockoff
				& 0.118 & 1.000
				& 0.251 & 1.000
				& 0.221 & 0.647
				& 0.293 & 0.627 \\
				
				$W_n$-Lasso
				& 0.203 & 1.000
				& 0.203 & 1.000
				& 0.269 & 1.000
				& 0.269 & 1.000 \\
				
				\midrule
				\multicolumn{9}{c}{$n_2=200$} \\
				\midrule
				FMCT-NN
				& 0.017 & 0.943
				& 0.030 & 0.965
				& 0.014 & 1.000
				& 0.023 & 1.000 \\
				
				FMCT-Lasso
				& 0.028 & 0.978
				& 0.084 & 1.000
				& 0.005 & 0.853
				& 0.040 & 0.927 \\
				
				Knockoff
				& 0.136 & 1.000
				& 0.255 & 1.000
				& 0.168 & 0.610
				& 0.223 & 0.600 \\
				
				$W_n$-Lasso
				& 0.213 & 1.000
				& 0.214 & 1.000
				& 0.232 & 1.000
				& 0.232 & 1.000 \\
				
				\bottomrule
			\end{tabular*}
			\label{table:FDR_E1_E6}
		\end{table}
	}

	\begin{example}
		Consider
		$$Y=5\sin(X_1)+5\sin(X_{20})+\exp(-2X_3)+0.1\epsilon,$$
		where $X\sim N_{20}(0,\Sigma)$ with $\Sigma_{rs}=0.5^{|r-s|}$ for $r,s=1,\ldots,20$, and $\epsilon\sim N(0,1)$ is independent of $X$.
		Thus, $X_1,X_3$, and $X_{20}$ are active.
	\end{example}
	
	Table~\ref{table:FDR_E1_E6} reports the empirical FDR and power. Across both Euclidean examples, the empirical FDR of both FMCT variants remains below the nominal level for every reported sample size. In Example~1, their power increases with $n_2$ and approaches one. In Example~6, FMCT-NN has high power throughout, whereas the power of FMCT-Lasso increases steadily with $n_2$. Model-X knockoffs have moderate power in Example~6 without a clear increase with $n_2$. The $W_n$-Lasso procedure attains power one throughout, but its empirical FDR, like that of knockoffs, is numerically above the corresponding nominal level in every reported setting.

	Due to space constraints, results for the three metric-space response settings (Example 3--5) appear in Section~\ref*{Appendix A} of the Supplementary Material.

	\FloatBarrier

	\section{Real data analysis}
	\label{sec:Real data analysis}
	We apply FMCT to hourly New York City taxi data \footnote{The New York City taxi data are available from
	\href{https://www1.nyc.gov/site/tlc/about/tlc-trip-record-data.page}{https://www1.nyc.gov/site/tlc/about/tlc-trip-record-data.page}} to identify predictor coordinates associated with network flow after adjustment for the remaining variables. The processed data consist of $N=n_1+n_2=1416$ hourly observations. Each hourly flow network is represented by a symmetric $3\times3$ weighted adjacency matrix, and its matrix exponential is used as the SPD response. The 14 predictors comprise nine hourly taxi service variables and five daily weather variables. Supplementary Material~\ref*{Appendix A} describes the data processing and spatial aggregation, and Table~\ref*{table:Variables in New York taxi data} lists the predictor coordinates.

	To emulate the semi-supervised design, we randomly select $n_1=1204$ observations whose response labels are masked and retain $n_2=212$ labeled observations, giving $n_1:n_2\approx17:3$. The Lasso regressions are fitted using the predictor-only sample, with the penalty selected by 10-fold cross-validation. We use the Log-Cholesky distance for the SPD response, with the remaining implementation settings as in the simulations. We conduct the coordinatewise tests at level $0.05$ and simultaneous testing at FDR levels $0.05$ and $0.1$.
	
	As shown in Table~\ref{table:real data results}, at level $0.05$, the coordinatewise tests reject the marginal coordinate null hypotheses for Ave. Distance ($X_1$), Ave. Fare ($X_2$), Ave. Passengers ($X_3$), Ave. Tip ($X_4$), Cash ($X_5$), Free ($X_8$), and Late Hour ($X_9$).
	Under simultaneous testing, only Ave. Fare ($X_2$) and Cash ($X_5$) are selected at both FDR levels. No weather variable is selected; this may partly reflect the coarser daily resolution of these predictors and does not establish that weather is conditionally irrelevant to taxi flows. Among the taxi service variables, fare and cash yield the strongest reported conditional evidence after adjustment for the remaining predictors, but the analysis does not support a causal interpretation. Because the primary FMCT uses $g(x)=x$, it may also miss contributions expressed only through conditional variances or other nonlinear features.
	A richer transformation family can be used to assess sensitivity to such structures.
	
	{\spacingset{1}
		\begin{table}[!htbp]
			\renewcommand{\arraystretch}{1.15}
			\centering
			\begin{threeparttable}
				\caption{Coordinatewise and simultaneous results for the New York City taxi data.}
				\begin{tabular}{c c c c c c c c}
					\hline
					Coordinate & $X_1$ & $X_2$ & $X_3$ & $X_4$ & $X_5$ & $X_6$ & $X_7$ \\
					\hline
					p-value & 0.016 & 0.000 & 0.050 & 0.038 & 0.000 & 0.182 & 0.248 \\
					e-value ($\alpha=0.05$) & 25.00 & 280.00 & 14.14 & 16.22 & 280.00 & 7.41 & 6.35 \\
					e-value ($\alpha=0.1$) & 17.68 & 140.00 & 10.00 & 11.47 & 140.00 & 5.24 & 4.49 \\
					Decision & $\ast$ & $\ast\!\ast\!\ast$ & $\ast$ & $\ast$ & $\ast\!\ast\!\ast$ &  &  \\
					\hline
					Coordinate & $X_8$ & $X_9$ & $X_{10}$ & $X_{11}$ & $X_{12}$ & $X_{13}$ & $X_{14}$ \\
					\hline
					p-value & 0.034 & 0.018 & 0.094 & 0.340 & 0.206 & 0.104 & 0.130 \\
					e-value ($\alpha=0.05$) & 17.15 & 23.57 & 10.31 & 5.42 & 6.97 & 9.81 & 8.77 \\
					e-value ($\alpha=0.1$) & 12.13 & 16.67 & 7.29 & 3.83 & 4.92 & 6.93 & 6.20 \\
					Decision & $\ast$ & $\ast$ &  &  &  &  &   \\
					\hline
				\end{tabular}
				\label{table:real data results}
				\begin{tablenotes}
					\footnotesize
					\item[\#] Asterisks denote the strongest applicable decision: $\ast$ for coordinatewise rejection at level $0.05$, and $\ast\!\ast$ and $\ast\!\ast\!\ast$ for selection by the simultaneous procedure at FDR levels $0.1$ and $0.05$, respectively.
				\end{tablenotes}
			\end{threeparttable}
		\end{table}
	}

\section{Conclusion}
\label{Conclusion}
This paper develops FMCT for marginal coordinate testing with Euclidean predictors and a metric space response. The scientific null $H_{0j}:Y\indep X_j\mid X_{-j}$ implies the auxiliary conditional mean restriction targeted by the primary statistic. Thus, rejection provides evidence against $H_{0j}$, whereas nonrejection does not establish conditional independence. 
Under the stated conditions, we derive a weighted sum of centered chi-square variables as the null limit, establish the validity of the wild bootstrap and consistency against fixed detectable alternatives, characterize power under local mean-element alternatives, and obtain asymptotic FDR control through truncated p-to-e calibration and e-BH.
	
Prespecified finite transformation families can probe broader alternatives, with analogous guarantees for null calibration, bootstrap validity, power, and simultaneous inference under suitable conditions. Full active-set recovery additionally requires coordinate exhaustiveness. Future work may develop data-adaptive transformation families, establish uniform bootstrap and FDR guarantees for growing $p$.

	\bibliographystyle{plainnat}
	\bibliography{reference.bib}

\end{document}